\documentclass[12pt]{article}

\begin{document}

\baselineskip=18.6pt plus 0.2pt minus 0.1pt

 \def\be{\begin{equation}}
  \def\ee{\end{equation}}
  \def\bea{\begin{eqnarray}}
  \def\eea{\end{eqnarray}}
  \def\nn{\nonumber\\ }
\newcommand{\nc}{\newcommand}
\nc{\bib}{\bibitem}
\nc{\cp}{\C{\bf P}}
\nc{\la}{\lambda}
\nc{\C}{\mbox{\hspace{1.24mm}\rule{0.2mm}{2.5mm}\hspace{-2.7mm} C}}
\nc{\R}{\mbox{\hspace{.04mm}\rule{0.2mm}{2.8mm}\hspace{-1.5mm} R}}

\begin{titlepage}
\title{
\begin{flushright}
 {\normalsize \small
IFT-UAM/CSIC-04-05 }
 \\[1cm]
 \mbox{}
\end{flushright}
{\bf On toric geometry, $Spin(7)$ manifolds,}\\[.3cm]
{\bf and type II superstring compactifications}
\author{Adil Belhaj$^1$\thanks{{\tt adil.belhaj@uam.es}}\ \ and\ J{\o}rgen
Rasmussen$^2$\thanks{{\tt rasmusse@crm.umontreal.ca}}\\[.3cm]
{\it \small $^1$Instituto de F\'{\i}sica Te\'orica, C-XVI,
Universidad Aut\'onoma de Madrid} \\ {\it \small E-28049-Madrid,
Spain}
\\[.3cm]
{\it\small $^2$Centre de Recherches Math\'ematiques, Universit\'e
de Montr\'eal} \\ {\it\small C.P. 6128, succursale centre-ville,
Montr\'eal, PQ, Canada H3C 3J7} } } \maketitle
\thispagestyle{empty}
\begin{abstract}
We consider type II superstring compactifications on the
singular $Spin(7)$ manifold constructed as a cone on
$SU(3)/U(1)$.
Based on a toric realization of the projective space $\cp^2$,
we discuss how the manifold can be viewed as three intersecting
Calabi-Yau conifolds. The geometric transition
of the manifold is then addressed in this setting.
The construction is readily extended to
higher dimensions where we speculate on possible
higher-dimensional geometric
transitions. Armed with the toric description of the
$Spin(7)$ manifold, we discuss a brane/flux duality
in both type II superstring theories compactified on this
manifold.
\end{abstract}
{\tt  KEYWORDS}: Toric geometry, superstrings,
  $Spin(7)$ manifold, compactification, geometric transition.

\end{titlepage}

\newpage

\section{Introduction}

Calabi-Yau conifold transitions in superstring compactifications
have been studied intensively over the last couple of years.
These transitions have become a 'standard' tool in understanding large
$N$ dualities. An example of such a duality is the equivalence of the $SU(N)$
Chern-Simons theory on ${\bf S}^3$ for large $N$, and the closed
topological strings on the resolved conifold \cite{GV}.
A further example has been obtained by embedding these results
in type IIA superstring theory \cite{V}.
In particular, the scenario with $N$ D6-branes wrapped around
the ${\bf S}^3$ of the deformed conifold $T^\ast {\bf S}^3$ for large $N$
has been found to be equivalent to type IIA superstrings on
the resolved conifold with
$N$ units of R-R two-form fluxes through ${\bf S}^2$.
The latter thus gives the strong-coupling description of the
weak-coupling physics of the former \cite{GV}.
This result has been 'lifted' to the eleven-dimensional
M-theory \cite{AMV} where it corresponds to a so-called
flop duality in M-theory compactified on a manifold
with $G_2$ holonomy, for short, a $G_2$ manifold.

Quite recently, similar studies have been done in three
dimensions using either type IIA superstring compactification
on a $G_2$ manifold \cite{AW,GST}, or M-theory compactified
on a $Spin(7)$ manifold \cite{GST}. This $Spin(7)$
manifold is constructed as a cone on $SU(3)/U(1)$.
Upon reduction of the M-theory case to ten dimensions,
the original geometric transition involving a collapsing
${\bf S}^5$ and a growing $\cp^2$ may be interpreted
as a transition between two phases described by wrapped
D6-branes or R-R fluxes, respectively.

An objective of the present work is to continue the study
of geometric transitions and brane/flux dualities in
lower dimensions. We shall thus consider type II
superstrings propagating on the same $Spin(7)$ manifold
as above. By comparison with the known results for
the Calabi-Yau conifold transition, in particular,
we conjecture new brane/flux dualities in two dimensions.
The type IIA and type IIB superstrings are treated separately,
and we find that the resulting gauge theories in two dimensions
have only one supercharge each, so that ${\cal N}=1/2$
in both cases.

The present study utilizes a toric geometry description of the
$Spin(7)$ manifold. We find that the manifold
can be viewed as three
intersecting Calabi-Yau conifolds associated to a triangular
toric diagram.
This result offers a picture for understanding the
topology-changing transition of the $Spin(7)$ manifold.
It also allows us to discuss the aforementioned brane/flux
transition based on an analysis of type II
superstrings on the individual Calabi-Yau conifolds.

Our toric description of the $Spin(7)$ manifold may
be extended to higher-dimensional manifolds thus
suggesting that (generalized) geometric transitions
may play a role in higher dimensions as well.
We propose an explicit hierarchy of pairs of
geometries related by such transitions.

The remaining part of this paper is organized as follows.
In Section 2, we use toric geometry to discuss the Calabi-Yau
conifold transition and its extension to the $Spin(7)$
manifold, and speculate on a further generalization to higher
dimensions. We then turn to compactifications of
superstrings in Section 3. Since our analysis is based
on the toric description of the $Spin(7)$ manifold, our results are
deduced from similar results on compactifications  on
Calabi-Yau conifolds. The associated brane/flux dualities
are discussed separately for type IIA and type IIB superstring
propagations. Section 4 contains some concluding remarks.

\section{Toric geometry and geometric transitions}

\subsection{Projective spaces and odd-dimensional spheres}

As a description of projective spaces in terms of toric geometry
lies at the heart of our study of superstring compactifications,
we shall review it here. Odd-dimensional (real) spheres are
equally important in our analysis and are therefore also discussed here.

The simplest (complex) projective space is $\cp^1$ with
a toric $U(1)$ action having two fixed points, $v_1$ and $v_2$,
corresponding to the North and South poles, respectively,
of the (real) two-sphere ${\bf S}^2\sim\cp^1$. In this way, $\cp^1$
may be viewed as the interval $[v_1,v_2]$
\be
      \mbox{
         \begin{picture}(20,20)(40,0)
        \unitlength=2cm
        \thicklines
    \put(0,0){\line(1,0){2}}
     \put(-0.25,0){$v_1$}
     \put(2.1,0){$v_2$}
  \end{picture}
}
\label{line}
\ee
referred to as the toric diagram,
with a circle on top which vanishes at the end points $v_1$
and $v_2$.

Embedded in $\C^3$, $\cp^2$ may be described as the
space of three complex numbers $(z_1,z_2,z_3)$ not all zero,
modulo the identification $(z_1,z_2,z_3)\sim(\la z_1,\la z_2,\la z_3)$
for all non-zero $\la\in\C$. Alternatively, $\cp^2$ is the (complex)
two-dimensional space with a toric $U(1)^2$ action with three
fixed points, $v_1$, $v_2$ and $v_3$. Its toric diagram
is the triangle $(v_1 v_2 v_3)$
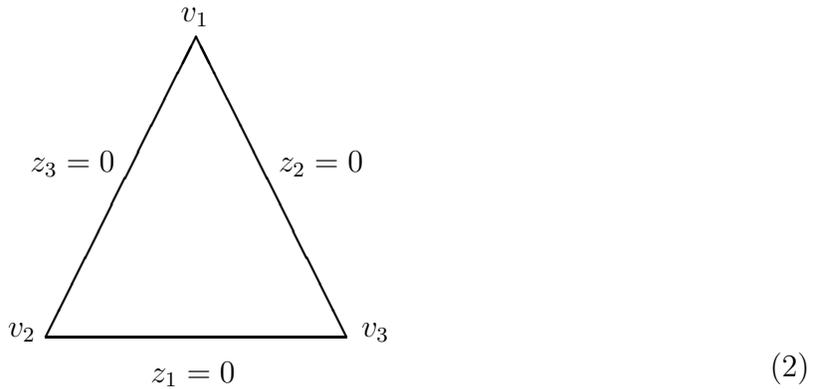
\be
\mbox{
 \begin{picture}(100,150)(0,-15)
    \unitlength=2cm
  \thicklines
   \put(0,0){\line(1,2){1}}
   \put(0,0){\line(1,0){2}}
  \put(2,0){\line(-1,2){1}}
 \put(0.9,2.1){$ v_1$}
 \put(-0.25,0){$v_2$}
   \put(2.1,0){$v_3$}
   \put(-0.1,1.1){$z_3=0$}
  \put(0.7,-0.3){$z_1=0$}
   \put(1.55,1.1){$z_2=0$}
\end{picture}
}
\label{triangle}
\ee
describing the intersection of three $\cp^1$'s.
Each of the three edges, $[v_1,v_2]$, $[v_2,v_3]$ and $[v_3,v_1]$,
is characterized by the vanishing of one of the homogeneous
coordinates: $z_3=0$, $z_1=0$ or $z_2=0$, respectively.
Each edge is stable under the action of a subgroup of
$U(1)^2$ -- two of them being the two $U(1)$ factors, while
the third subgroup is the diagonal one.
This toric realization of $\cp^2$ can be viewed as the
triangle $(v_1v_2v_3)$ with a torus, ${\bf T}^2$, on top
which collapses to a circle at an edge and to a point at
a vertex.

This representation is readily extended to the $n$-dimensional
projective space $\cp^n$ where we have a ${\bf T}^n$ fibration over
an $n$-dimensional simplex (regular polytope), see \cite{LV}, 
for example.
In this case, the ${\bf T}^n$ collapses
to a ${\bf T}^{n-1}$ on each of the $n$ faces of the simplex,
and to a ${\bf T}^{n-2}$ on each of the $(n-2)$-dimensional intersections
of these faces, etc. We recall that $\cp^n$ is defined
similarly to $\cp^2$ in terms of $n+1$ homogeneous coordinates
modulo the identification $(z_1,...,z_{n+1})\sim(\la z_1,...,\la z_{n+1})$.

The odd-dimensional (real) spheres admit a similar description.
The one-sphere, for example, is trivially realized as a
${\bf T}^1\sim{\bf S}^1$ over the zero-simplex -- a point. The three-sphere
may be realized as a ${\bf T}^2$ over a one-simplex -- a line segment
as the one in (\ref{line}). This may be extended to the
$(2n+1)$-dimensional sphere ${\bf S}^{2n+1}$ which may
be described as a ${\bf T}^{n+1}$ over an $n$-simplex.
Of particular interest is the five-sphere ${\bf S}^5$ which
in this way may be realized as the triangle (\ref{triangle})
with a ${\bf T}^3$ on top (whereas $\cp^2$ had a ${\bf T}^2$
on top). It is stressed that it is for $n=1$ only that the even-dimensional
sphere ${\bf S}^{2n}$ is equivalent to $\cp^n$.

To illustrate this toric description of odd-dimensional spheres,
let us add a couple of comments on ${\bf S}^5$ realized as
a ${\bf T}^3$ over a triangle. As in (\ref{triangle}),
an edge of the triangle corresponds to the vanishing of one
of the three complex coordinates of the embedding space
$\C^3$. Each edge of the triangle (\ref{triangle}) is stable under
the action of one of the three $U(1)$ factors of $U(1)^3$
associated to ${\bf T}^3$. The three-torus itself collapses to
a two-torus ${\bf T}^2$ at an edge and to a circle at a vertex.
We may thus view ${\bf S}^5$ as three intersecting three-spheres
over the triangle (\ref{triangle}).
As opposed to the toric description of $\cp^2$ as a ${\bf T}^2$
over a triangle, the diagonal $U(1)$ of $U(1)^3$ in the ${\bf S}^5$
description has no fixed points.
This is natural from the realization of $\cp^2$ as
${\bf S}^5$ modulo $U(1)$.

\subsection{Calabi-Yau conifold}

We shall also make use of the non-compact
Calabi-Yau threefold defined in $\C^4$ by the equation
\be
 uv-xy\ =\ 0\ .
\label{uvxy}
\ee
It may be viewed as the singular cone on the five-dimensional
base ${\bf S}^2\times{\bf S}^3$ and is therefore referred
to as the Calabi-Yau conifold.
The singularity is located at the origin and may be turned into
a regular point by blowing it up. There are basically two ways of doing
that, referred to as resolution and deformation, respectively.
Resolving the singularity consists in replacing the  singular point
by a $\cp^1$. In this way, the local geometry is given by an
$O(-1)+O(-1)$ bundle over $\cp^1$. The smooth manifold thus obtained
is called the resolved conifold and is of topology $\R^4\times \cp^1$.
In the case of complex deformation, the conifold singularity
is removed by modifying the defining algebraic equation
(\ref{uvxy}) by introducing the complex parameter $\mu$:
\be
 uv-xy\ =\ \mu\ ,
\label{uvxymu}
\ee
 while keeping the K\"ahler structure. The origin is thereby
 replaced by ${\bf S}^3$, and the local geometry is given by
 $T^\ast{\bf S}^3$ of topology $\R^3\times{\bf S}^3$.
 This is called the deformed conifold and is related to the
 resolved conifold by the so-called conifold transition.

This conifold transition admits a representation
in toric geometry, where it can be understood as an
enhancement or breaking, respectively, of the toric
circle actions. On the one hand, the $O(-1)+O(-1)$ bundle
over $\cp^1$ has only one toric $U(1)$ action, identified
with the toric action on $\cp^1$ itself, while the deformed
conifold $T^\ast{\bf S}^3$ has a toric $U(1)^2$
action since the spherical part can be viewed as a
${\bf T}^2$ over a line segment.
Referring to (\ref{uvxymu}), the torus is generated
by the two $U(1)$ actions
\be
 (u,v)\ \rightarrow\ (e^{i\theta_1}u,e^{-i\theta_1}v)\ ,\ \ \ \ \ \ \ \
 (x,y)\ \rightarrow\ (e^{i\theta_2}x,e^{-i\theta_2}y)
\ee
with $\theta_i$ real.
Thus, the blown-up ${\bf S}^3$ may be described by the
complex interval $[0,\mu]$ with the two circles parameterized
by $\theta_i$ on top, where ${\bf S}^1(\theta_1)$
collapses to a point at $\mu$ while ${\bf S}^1(\theta_2)$
collapses to a point at $0$.
The transition occurs when one of these
circles refrains from collapsing while the other one collapses
at both interval endpoints. This breaks the toric
$U(1)^2$ action to $U(1)$, and the missing $U(1)$ symmetry
has become a real line (over $\cp^1$). The resulting geometry
is thus the resolved conifold. The following picture may help
to illustrate this transition which can go in both directions:
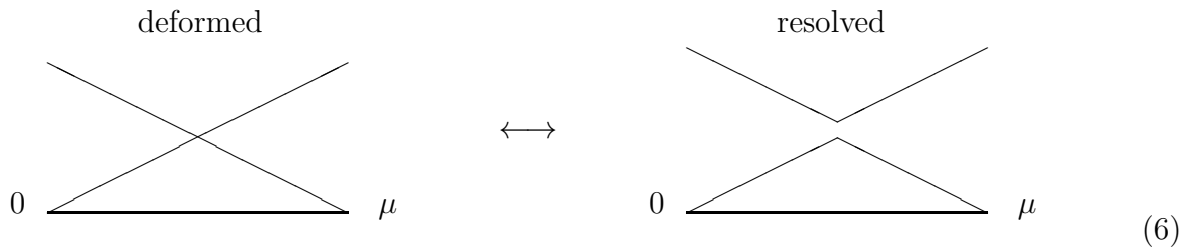
\be
      \mbox{
         \begin{picture}(100,110)(140,-10)
        \unitlength=2cm
        \thicklines
    \put(0,0){\line(1,0){2}}
     \put(-0.25,0){$0$}
     \put(2.2,0){$\mu$}
 \thinlines
  \put(0,0){\line(2,1){2}}
  \put(2,0){\line(-2,1){2}}
   \put(4.25,0){\line(2,1){1}}
  \put(6.25,0){\line(-2,1){1}}
  \put(4.25,1.1){\line(2,-1){1}}
  \put(6.25,1.1){\line(-2,-1){1}}
  \thicklines
  \put(4.25,0){\line(1,0){2}}
     \put(4,0){$0$}
     \put(6.45,0){$\mu$}
  \put(3,0.5){$\longleftrightarrow$}
  \put(4.85,1.2){resolved}
  \put(0.6,1.2){deformed}
 \end{picture}
}
\label{gra}
\ee
The top, thin and piecewise straight
line in the resolved part of (\ref{gra}) corresponds
to the extra $\R$ while the remaining three thin lines
(the two straight lines in the deformed part, and the lower,
thin and piecewise straight line segment in the resolved part)
indicate the $U(1)$'s. The two thick line segments represent the
underlying interval.

A somewhat pragmatic way of viewing the conifold
transition is based on the conical structure of the conifold
itself as a cone on ${\bf S}^2\times{\bf S}^3$.
As described in \cite{AW}, an $n$-dimensional
cone on an $(n-1)$-dimensional compact space $Y$
with metric $d\Omega^2$ has metric
\be
 ds^2\ =\ dr^2+r^2d\Omega^2\ .
\label{met}
\ee
It has a singularity at the origin unless $Y={\bf S}^{n-1}$
and $d\Omega^2$ is the standard 'round' metric.
In that case the cone corresponds to $\R^n$.
Now, the deformed conifold is obtained by 'pulling'
the conical structure off of the ${\bf S}^3$ factor in the base,
while 'maintaining' it on the ${\bf S}^2$ factor.
The latter is then equivalent to $\R^3$ and we have
recovered the $\R^3\times{\bf S}^3$ structure of the
deformed conifold. The resolved conifold is obtained in
a similar way by pulling off the conical structure of the
${\bf S}^2$ factor.

\subsection{$Spin(7)$ manifolds}

Here we shall present a picture for understanding the
topology-changing geometric transition of the $Spin(7)$
manifold discussed in \cite{GST} and alluded to in
Section 1. First we recall that
a $Spin(7)$ manifold is a real eight-dimensional
Riemannian manifold with holonomy group $Spin(7)$.
As in the case of Calabi-Yau and $G_2$ manifolds, there
are several such geometries \cite{J}.
The example we shall be interested in may be described
as a singular real cone over the seven-dimensional
Aloff-Wallach (coset) space $SU(3)/U(1)$.
It was argued in \cite{GST} that there are two ways of
blowing up the singularity, replacing the singularity
by either $\cp^2$ or ${\bf S}^5$. The resulting
smooth $Spin(7)$ manifolds have topologies
\be
 \mbox{resolution}:\ \ \ \ \ \ \ Spin(7):\ \ \R^4\ \times\ \cp^2\ \ \
  \ \ \ \ (\mbox{Calabi-Yau}:\ \ \R^4\ \times\ \cp^1)
\label{res}
\ee
and
\be
 \mbox{deformation}:\ \ \ \ \ \ Spin(7):\ \ \R^3\ \times\ {\bf S}^5\ \ \
  \ \ \ \ (\mbox{Calabi-Yau}:\ \ \R^3\ \times\ {\bf S}^3)\ ,
\label{def}
\ee
and are referred to as resolution and deformation, respectively,
due to the similarity with the Calabi-Yau conifold discussed above
(and indicated in (\ref{res}) and (\ref{def})).

Our aim here is to re-address the transition between these two
manifolds using toric geometry. As described in the following,
the basic idea is to view
the singular $Spin(7)$ manifold (the real cone on $SU(3)/U(1)$)
as three intersecting Calabi-Yau conifolds associated to the
triangular toric diagram (\ref{triangle}). To reach this picture,
we first recall that a deformed $Spin(7)$ manifold is obtained by 
blowing up an ${\bf S}^5$, while a resolved $Spin(7)$ manifold is
obtained by blowing up a $\cp^2$. Deformed and resolved
conifolds, on the other hand, are obtained by blowing up
an ${\bf S}^3$ or a $\cp^1$, respectively. Since an ${\bf S}^5$
may be represented by three intersecting three-spheres, while
$\cp^2$ may be represented by three intersecting two-spheres,
we thus see that the deformed and resolved $Spin(7)$
manifolds correspond to three intersecting
conifolds being deformed or resolved, respectively.

To recapitulate this, 
let us consider $\C^3$ parameterized by
$(z_1,z_2,z_3)$. A five-sphere is obtained
by imposing the constraint
\be
 |z_1|^2+|z_2|^2+|z_3|^2\ =\ r
\label{r}
\ee
(with $r$ real and positive), while the additional identification
\be
 (z_1,z_2,z_3)\ \sim\ (e^{i\theta}z_1,e^{i\theta}z_2,e^{i\theta}z_3)
\label{theta}
\ee
(with $\theta$ real)
will turn it into a $\cp^2$. In either case, $r$ measures the size.
With both conditions imposed, we can obtain the
three resolved Calabi-Yau conifolds
\be
 \R^4\ \times\ \cp^1(z_k=0)\ ,\ \ \ \ \ \ \ \ k=1,2,3
\label{res123}
\ee
embedded in $\R^4\times\C^3$,
simply by setting one of the coordinates equal to 0.
With reference to the triangle (\ref{triangle}), this means that 
the resolution of the $Spin(7)$ singularity reached by blowing up
a $\cp^2$ may be described by three intersecting
resolved conifolds over the triangle (\ref{triangle}).
Likewise, the deformation of the $Spin(7)$ singularity
constructed by blowing up an ${\bf S}^5$ may be
realized as three intersecting deformed Calabi-Yau
conifolds
\be
 \R^3\ \times\ {\bf S}^3(z_k=0)\ ,\ \ \ \ \ \ \ \ k=1,2,3
\label{def123}
\ee
over the same triangle. This description is thus based on
our toric representation of ${\bf S}^5$ as a ${\bf T}^3$ over a
triangle. As already mentioned, this construction collapses to a ${\bf T}^2$
over an edge for $z_k=0$ where $k=1$, 2 or 3, and
it is recalled that the resulting ${\bf T}^2$ over a line
segment corresponds to ${\bf S}^3$.

Since the basic intersection of the deformed or resolved
conifolds is governed by the constituent three- or two-spheres,
one may describe the intersection of the conifolds by
the intersection matrices associated to ${\bf S}^5$ or $\cp^2$,
respectively. They read
\bea
 M_{def}\ =\ \left(\begin{array}{ccc} 0&1&1\\ \\
      1&0&1 \\ \\ 1&1&0\end{array}\right),\ \ \ \ \ \ \ 
 M_{res}\ =\ \left(\begin{array}{ccc} -2&1&1\\  \\
      1&-2&1 \\  \\ 1&1&-2\end{array}\right)
\eea

We emphasize that the $Spin(7)$ transition in our
picture is accompanied
by Calabi-Yau conifold transitions. In the transition
from the resolved $Spin(7)$ manifold (\ref{res})
to the deformed one (\ref{def}), for example,
this indicates that the collapsing $\cp^2$ and its constituent
two-spheres are replaced by ${\bf S}^5$ and
its constituent three-spheres. This is at the core of the
dualities in the phase transition of the compactified
superstrings to be discussed below. 

One may attempt
to illustrate the geometric transitions of $Spin(7)$ manifolds
by generalizing (\ref{gra}).
The following proposal extends readily to higher dimensions (see below), 
and reads
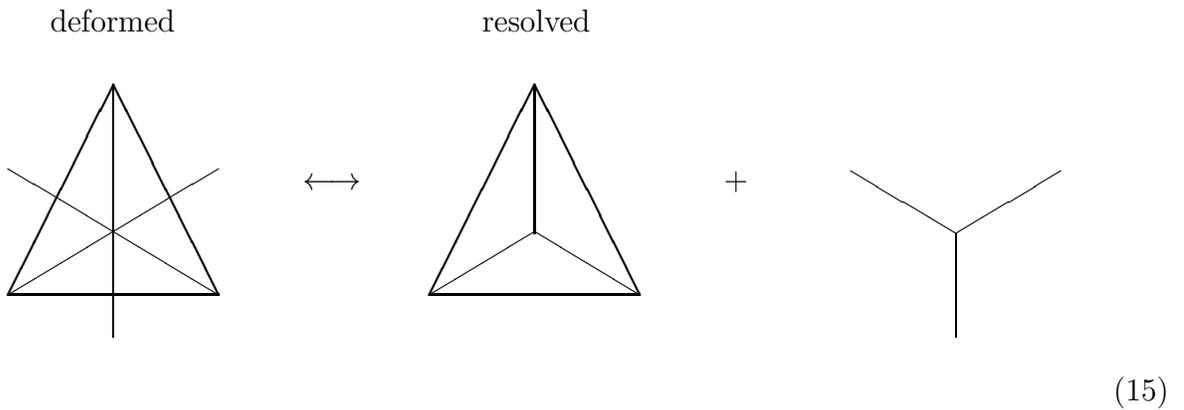
\be
 \mbox{
 \begin{picture}(100,160)(150,-40)
    \unitlength=1.4cm
  \thicklines
   \put(0,0){\line(1,2){1}}
   \put(0,0){\line(1,0){2}}
  \put(2,0){\line(-1,2){1}}
  \put(0.4,2.5){deformed}
 \thinlines
  \put(0,0){\line(5,3){2}}
  \put(2,0){\line(-5,3){2}}
  \put(1,2){\line(0,-1){2.4}}
 \thicklines
   \put(4,0){\line(1,2){1}}
   \put(4,0){\line(1,0){2}}
  \put(6,0){\line(-1,2){1}}
  \put(4.5,2.5){resolved}
  \put(2.8,1){$\longleftrightarrow$}
  \put(6.8,1){$+$}
 \thinlines
  \put(4,0){\line(5,3){1}}
  \put(6,0){\line(-5,3){1}}
  \put(5,2){\line(0,-1){1.42}}
  \put(9,0.58){\line(5,3){1}}
  \put(9,0.58){\line(-5,3){1}}
  \put(9,0.58){\line(0,-1){0.98}}
\end{picture}
}
\label{triangle2}
\ee
The configuration to the left is a representation of the ${\bf S}^5$ part
of the deformed $Spin(7)$ manifold, in which the three thin lines
represent the three $U(1)$ factors. The counting is less obvious to the
right of the arrow, as the triangular part represents the $\cp^2$ part
of the resolved $Spin(7)$ manifold. The inscribed three-vertex in thin 
line segments
thus corresponds to two $U(1)$ factors only. The three-vertex to the
far right represents the extra $\R$ in the resolved scenario (\ref{res}).

The $Spin(7)$ transition may be viewed as taking place
when passing a
particular point while moving along a particular curve in
the moduli space of $Spin(7)$ manifolds.
The point corresponds to the singular $Spin(7)$
manifold, whereas the remaining points on the curve
are associated to the deformed and resolved
$Spin(7)$ manifolds. In one direction away from the
singular point, the points correspond
to the deformed manifolds with the size $r$ of the
blown-up ${\bf S}^5$ parameterizing that part of
the curve. Likewise, the other direction away from the
singular point is parameterized by the
size $r$ of the blown-up $\cp^2$ in the resolved
$Spin(7)$ manifold. The singular point is 'shared'
as it is reached from either side when the relevant
$r$ vanishes: $r=0$.

In the interpretation of the $Spin(7)$ manifold as
three intersecting Calabi-Yau manifolds over a triangle,
we see that the $Spin(7)$ transition corresponds to all three
Calabi-Yau manifolds undergoing simultaneous
conifold transitions. We find it an interesting problem to
understand the geometries associated to individual
conifold transitions and hope to report on it elsewhere.
Our graphic representation (\ref{triangle2}) (and (\ref{gra})) 
does not seem to shed light on this as it is 
based on the transition of the full blow-ups, i.e., ${\bf S}^5$ and $\cp^2$,
and not on their constituent three- and two-spheres.

\subsection{On possible extensions}

It seems possible to extend our previous analysis of the
$Spin(7)$ manifold in terms of intersecting Calabi-Yau manifolds
to higher dimensions.
To this end, let us consider the complex $(n+1)$-dimensional space
$\C^{n+1}$ parameterized by
$(z_1,...,z_{n+1})$. A $(2n+1)$-dimensional sphere is obtained
by imposing the constraint
\be
 \sum_{j=1}^{n+1}|z_j|^2\ =\ r
\label{rgen}
\ee
(with $r$ real and positive), while the additional identification
\be
 (z_1,...,z_{n+1})\ \sim\ (e^{i\theta}z_1,...,e^{i\theta}z_{n+1})
\label{thetagen}
\ee
(with $\theta$ real)
will turn it into $\cp^n$. In either case, $r$ measures the size of
the resulting space.
Since ${\bf S}^{2n+1}$ can be described as a ${\bf T}^{n+1}$
over an $n$-simplex it supports a toric $U(1)^{n+1}$ action
whereas $\cp^n$ (which may be realized as a ${\bf T}^n$
over an $n$-simplex) admits a toric $U(1)^n$ action.
The additional $U(1)$ is the one used in the identification
(\ref{thetagen}). As in the picture (\ref{gra}), we are thus
expecting that a geometric transition can take place,
replacing a $U(1)$ by the one-dimensional real line $\R$.
Since the $U(1)$ is associated to one of the ${\bf S}^1$
factors of ${\bf T}^{n+1}$, the transition essentially amounts to
replacing ${\bf T}^{n+1}$ by ${\bf T}^n\times\R$.
Our interest is in real fibrations over the
spaces ${\bf S}^{2n+1}$ and $\cp^n$
so the relevant geometric transitions would read
\be
 (\mbox{deformed})\ \ \ \ \ \ \R^m\ \times\ {\bf S}^{2n+1}\ \ \ \ 
  \longleftrightarrow\ \ \ \
 \R^{m+1}\ \times\ \cp^n\ \ \ \ \ \ (\mbox{resolved})\ .
\label{m}
\ee
With $m=3$, we expect to be able to
describe the transition (\ref{m}) in terms
of Calabi-Yau conifolds. Using arguments similar to
the $Spin(7)$ example above, this generalized geometric
transition should be related to $\frac{1}{2}n(n+1)$
intersecting conifolds over the $n$-simplex,
where the number of conifolds is equal to the number
of one-dimensional edges of the simplex.
One should also expect to be able
to describe the transition in terms of $\frac{1}{6}n(n^2-1)$
intersecting $Spin(7)$ manifolds over the $n$-simplex,
where the number of them is equal to the number of two-dimensional
faces of the simplex.
We hope to address this further elsewhere.

The extension of the transition picture (\ref{triangle2}) 
to higher dimensions is based on $\cp^n$ and 
${\bf S}^{2n+1}$ admitting descriptions as ${\bf T}^n$
and ${\bf T}^{n+1}$ fibrations, respectively, 
over an $n$-simplex. One chooses an extra point different from
the $n+1$ vertices of the $n$-simplex
in such a way that any subset of $q<n+2$ nodes out
of the total of $n+2$ points gives
rise to a $(q-1)$-simplex. A natural choice is the centre of the
original (regular) $n$-simplex. To represent ${\bf S}^{2n+1}$, 
one then draws thin lines from the vertices
through this extra point, where the thin lines represent the
$U(1)$ factors of $U(1)^{n+1}$, cf. (\ref{triangle2}). 
The projective counterpart, $\cp^n$, is
represented by ending these thin lines at the common point, 
resulting in an $(n+1)$-vertex inscribed in the $n$-simplex. 
This inscribed vertex corresponds to $U(1)^n$.
Finally, the real line $\R$ may be represented by
a 'free' $(n+1)$-vertex, and the graphical representation
of the transition (\ref{m})
is a higher-dimensional version of (\ref{triangle2}).

The conifold analysis based on the toric
variety $\cp^2$ could alternatively be extended by blowing up some generic
points. With the number
of points restricted as $k=1,2,3$, this defines a so-called
toric del Pezzo surface denoted $dP_k$.
The blowing up consists in replacing a point by $\cp^1$ with
a line segment as its toric diagram. The full del Pezzo surface
will thus have a polygon with $k+3$ legs as its toric diagram.

\section{Type II superstring compactifications}

\subsection{Compactification on conifold}

Based on the Calabi-Yau conifold transition discussed above,
Gopakumar and Vafa have argued that the $SU(N)$ Chern-Simons
theory on ${\bf S}^3$ for large $N$
is dual to topological strings on the resolved conifold \cite{GV}.
In this way, the 't Hooft expansion of the Chern-Simons free energy
has been shown to be in agreement, for all genera, with the
topological string amplitudes on the resolved conifold.
This duality has subsequently been embedded in type IIA
superstring theory \cite{V}, where it was proposed that $N$
D6-branes wrapped around the three-sphere of the deformed
conifold is equivalent (for large $N$) to type IIA superstrings
on the resolved conifold with the D6-branes replaced by
$N$ units of R-R two-form fluxes through the two-sphere (${\bf S}^2\sim\cp^1$)
in the resolved conifold. This duality thus offers a way of understanding
the same physics at strong coupling.

The mirror version in type IIB superstring theory of this duality
states that the scenario with
$N$ D5-branes wrapped around the two-sphere in
the resolved conifold, is equivalent
(for large $N$) to three-form fluxes through the
${\bf S}^3$ of the deformed conifold.
This has been generalized to other Calabi-Yau
threefolds where the blown-up geometries
involve several $\cp^1$'s \cite{SV,CIV,CKV,CFIKV,AgMV,G,LL}.

The large $N$ duality in type IIA superstring theory has also
been lifted to M-theory \cite{AMV,AW} (see also \cite{Ach})
where it is known
to give a so-called flop duality. Unlike the duality in string
theory, the phase transition here is smooth and does not
correspond to a topology-changing geometric transition.

\subsection{Compactification on $Spin(7)$ manifold and brane/flux duality}

Based on the results on superstrings compactified on
Calabi-Yau threefolds and our toric description of
the geometric transition of the $Spin(7)$ manifolds,
we now consider the two-dimensional gauge theories
obtained by compactifying type II superstrings on
these $Spin(7)$ manifolds. The idea is to study the
consequences of adding $N$ wrapped D-branes
to the set-up before letting the manifold undergo
the geometric transition.
In the transition from the resolved to the deformed
$Spin(7)$ manifold, we initially have D-branes wrapping
$\cp^2$ (and its constituent two-spheres). We conjecture
that they are replaced, under the transition,
by R-R fluxes through ${\bf S}^5$ (and its constituent
three-spheres). Similarly in the transition from deformed
to resolved $Spin(7)$ manifolds, we conjecture that
D-branes wrapped around ${\bf S}^5$ (and its
constituent three-spheres) are replaced by R-R fluxes
through $\cp^2$ (and its constituent two-spheres).
The kind of D-branes involved and the more detailed
phase transition depend on which type II superstrings
are propagating on the $Spin(7)$ manifolds.
In the following we shall therefore consider type IIA and type IIB
separately. We find that they lead to different brane/flux
dualities.
\\[.2cm]
\noindent\underline{Duality in type IIB superstring theory.}\\
We start by considering type IIB superstrings on the resolved
$Spin(7)$ manifold (\ref{res}). Since the type IIB theory does
not support four-forms, one considers D5-branes wrapped around $\cp^2$.
A two-dimensional $U(N)$ gauge model can be obtained
by wrapping $N$ D5-branes on $\cp^2$. The volume of
$\cp^2$ described by $r$ (\ref{r}) is proportional to the
inverse of the gauge coupling squared. This two-dimensional
model has only one supercharge so ${\cal N}=1/2$.
Now, when the manifold undergoes the geometric transition
to the deformed $Spin(7)$ manifold (\ref{def}), the $N$ D5-branes
disappear and we expect a dual physics with $N$ units of
R-R three-fluxes through the compact three-cycles, ${\bf S}^3$,
in the intersecting Calabi-Yau threefolds .
These fluxes could be accompanied by some NS-NS fluxes
through the non-compact dual three-cycles in the six-dimensional
deformed conifolds. In order to handle the associated divergent 
integrals, one would have to introduce a cut-off to regulate
the infinity \cite{CIV}.
\\[.2cm]
\noindent\underline{Duality in type IIA superstring theory.}\\
Here we start with type IIA superstrings on the deformed
$Spin(7)$ manifold (\ref{def}). In this case, a two-dimensional
$U(N)$ gauge theory can be obtained by wrapping $N$
D6-branes around ${\bf S}^5$. As above, this gauge model
has only one supercharge, so again ${\cal N}=1/2$.
At the transition point, the D6-branes disappear and are
replaced by R-R two-form fluxes through the two-spheres
embedded in $\cp^2$ in the resolved $Spin(7)$ manifold
(\ref{res}).

One could wonder if there is an M-theory interpretation of this
type IIA transition. Let us therefore consider a nine-dimensional
manifold $X_9$ with a $U(1)$ isometry. M-theory compactified
on $X_9$ is then equivalent to type IIA superstrings compactified
on $X_9/U(1)$. We start with the resolved $Spin(7)$
manifold (\ref{res}) and identify the extra eleventh
compact dimension of M-theory with the ${\bf S}^1$
that generates (\ref{theta}). In this way, the extra M-theory
circle becomes the fiber in the definition of ${\bf S}^5$ as
an ${\bf S}^1$ fibration over $\cp^2$. We thus end up with
an $\R^4$ bundle over ${\bf S}^5$ as the compactification space
in M-theory. As a consequence, the moduli space of
M-theory on such a background is parameterized by the
the real parameter $r$ defining the volume of ${\bf S}^5$
(\ref{r}), and cannot be complexified by the C-field.
Starting with the resolved $Spin(7)$ manifold, on the other hand,
the eleventh M-theory dimension is obtained by extending
$\R^3$ to $\R^4$ with the isometry being
a trivial $U(1)$ action on the fiber $\R^4$.
Using arguments similar to those in \cite{AMV}, we conjecture
that this lift to M-theory gives rise to a (smooth) flop transition
in the $\R^4$ bundle over ${\bf S}^5$ where a five-sphere
collapses and is replaced by a five-sphere.
In our scenario, however, the physics resulting from the
type IIA superstring compactification
undergoes a singular phase transition.

\section{Discussion}

Based on toric geometry, we have studied geometric transitions
of $Spin(7)$ manifolds. Our framework allowed us to discuss
extensions to higher dimensions.
It also made it possible to address straightforwardly
type II superstring compactifications on $Spin(7)$ manifolds, from
which some brane/flux dualities were extrapolated.

Our work opens up for further studies. One interesting problem
is to understand better the geometries involved in our proposal
for higher-dimensional geometric transitions.
Another question is related to the toric description of the
$Spin(7)$ manifolds as intersecting Calabi-Yau threefolds
over a triangle where the $Spin(7)$ transition corresponds
to three simultaneous conifold transitions. A natural question
concerns the geometries
associated to individual conifold transitions. Of potential
importance to superstring and M-theory compactifications,
one should then study what the physical implications of such transitions
would be.
It would also be interesting to understand the link between our
results and the ones in \cite{GVW} based on string compactifications
on Calabi-Yau fourfolds. One approach to this problem could be to 
consider the $Spin(7)$ manifold as a Calabi-Yau fourfold modulo
an involution, thus ensuring the same number of supersymmetries.
We hope to report elsewhere on these open problems.
\\[.3cm]
{\bf Acknowledgments.}  AB would like to thank Cesar Gomez, Luis E. Ibanez,
Karl Landsteiner, Carlos Munoz, El Hassan Saidi and Angel Uranga 
for discussions and scientific help. 
AB is supported by Ministerio de Educaci\'{o}n Cultura y Deportes (Spain) 
grant SB 2002-0036.

\end{document}